\documentclass[runningheads,citeauthoryear]{apinv}
\usepackage{epsfig,cite,graphics}
\usepackage{marvosym} 

\usepackage[utf8]{inputenc}

\usepackage[colorlinks=true, citecolor=blue, linkcolor=blue, urlcolor=blue]{hyperref}
\usepackage{multirow}
\usepackage{enumitem}

\DeclareUnicodeCharacter{202F}{\,}
\DeclareUnicodeCharacter{2243}{\ensuremath{\simeq}}

\begin{document}

\title{Scaling Relations across Galaxy Classification Schemes: I. Star Formation Rate–Stellar Mass Plane of CALIFA Nearby Galaxies}
\titlerunning{Star Formation Rate–-Stellar Mass Plane of CALIFA Nearby Galaxies}
\author{Veselina Kalinova\inst{1,2}, Keiichi Kodaira\inst{3,4}, and Dario Colombo\inst{5}}
\authorrunning{V. Kalinova et al.}
\tocauthor{Veselina Kalinova et al.} 
\institute{Max Planck Institute for Radio Astronomy, Auf dem Hügel 69, 53121 Bonn, Germany
\and Institute of Astronomy and National Astronomical Observatory,
Bulgarian Academy of Sciences, 72 Tsarigradsko Chaussee Blvd., 1784 Sofia, Bulgaria
 \newline
	\email{kalinova@mpifr.de} 
     \and National Astronomical Observatory of Japan;  Osawa2-21-1, Mitaka-shi, Tokyo, Japan  PC 181-8588
         \and SOKENDAI, International Village, Hayama-machi, Miura-gun, Kanagawa-ken, Japan, PC 240-0193
         \and Argelander-Institut für Astronomie, University of Bonn, Auf dem Hügel 71, 53121 Bonn, Germany
    }
\papertype{Submitted on 13.11.2025; Accepted on 13.02.2026}	

\maketitle

\begin{abstract}
To gain deeper insights into galaxy evolution and the mechanisms driving transitions between different galaxy morphologies, we analyse the connection between star formation rate and stellar mass for 231 galaxies spanning Hubble types E1--Sdm from the Calar Alto Legacy Integral Field Spectroscopy Area survey using three complementary classification schemes. The Hubble classification provides structural information, the circular velocity curve classification$-$based on principal component analysis$-$ traces the total gravitational potential, and the Quenching classification$-$derived from H$_{\alpha}$ equivalent width maps$-$indicates the spatial extent of quenched regions relative to star-forming areas.
We find a clear separation of galaxy populations on the star formation rate--stellar mass plane. Late-type spirals with slow-rising circular velocity curves, represented by star-forming and quiescent-nuclear-ring galaxies, dominate the blue cloud. Early-type spirals with flat or round-peaked circular velocity curves belonging to centrally quiescent or mixed class populate the green valley, representing a transitional stage. Elliptical and lenticular galaxies with round- or sharp-peaked circular velocity curves from nearly retired or fully retired QSs reside on the red sequence. Furthermore, our results indicate that the morphological groups Sc--Scd, Sd--Sdm, and E1--E3 are characterized by a unique set of QSs and circular velocity curves, while galaxies with morphologies such as Sa--Sbc spread over multiple QSs and circular velocity curves. The distribution of the classification classes shows a tight link between galaxy structure, gravitational potential, and suppression of star-formation in the galaxies.
\end{abstract}


\section{Introduction}
\label{S:intro}
Scaling relations between galaxy properties provide fundamental constraints on the mechanisms that regulate galaxy evolution. Among these, the correlation between star formation rate (SFR) and stellar mass ($M_\ast$) is one of the most well-established (e.g., \cite{Brinchmann2004}, \cite{Daddi2007}, \cite{Speagle2014}, \cite{Cano-Diaz2016}, \cite{Sanchez2018}). This relation indicates that stellar mass is a primary predictor of star formation activity, with star-forming galaxies following a relatively tight sequence commonly referred to as the star-forming main sequence (MS). Deviations from the MS provide insight into galaxy evolution, with the scatter linked to variations in gas fraction, star formation efficiency, merger-driven bursts, or quenching processes (\cite{Rodighiero2011}, \cite{Schreiber2015}, \cite{Colombo2020}, \cite{Tacconi2020}, \cite{Colombo2025b}). Understanding how galaxies populate the MS—and what drives their departures—is therefore critical for constraining models of galaxy formation and evolution.

Recent studies (e.g., \cite{Baker2022}), on the other hand, suggest that the observed relation between SFR and $M_\ast$ may largely be a secondary correlation, primarily driven by the SFR–$M_{\rm mol}$ relation (where $M_{\rm mol}$ is the molecular gas mass) and the $M_{\rm mol}$–$M_\ast$ scaling. In this framework, $M_{\rm mol}$ acts as the key mediator: $M_\ast$ regulates the condensation of the interstellar medium, facilitating the transition from atomic to molecular gas, while star formation proceeds efficiently within these molecular clouds. Complementing this view, \cite{Tacchella2016} demonstrate using cosmological simulations, that galaxies move along and across the star-forming MS through recurrent phases of gas compaction, depletion, and replenishment, thereby regulating their star formation histories.

Nearby galaxies offer an ideal laboratory to study these processes in detail, as their stellar populations, gas content, and internal kinematics can be measured with high spatial and spectral resolution. Large spectroscopic integral-field surveys,  such as CALIFA (Calar Alto Legacy Integral Field spectroscopy Area; \cite{Sanchez2012}), SAMI (The Sydney-AAO Multi-object Integral field spectrograph; \cite{Croom2012}), and MaNGA (Mapping Nearby Galaxies at Apache Point Observatory; \cite{Bundy2015}), have enabled resolved and robust estimates of $M_\ast$, SFR, and kinematic structure across diverse galaxy populations. These datasets allow for a comprehensive examination of how structural and dynamical properties relate to star formation activity and galaxy evolution.

Previous studies have highlighted the interplay between galaxy morphology, central mass concentration, and star formation quenching. For example, high bulge-to-total ratios have been linked to reduced SFRs, consistent with morphological quenching scenarios (\cite{Martig2009}, \cite{Bluck2014}, \cite{Bluck2019}, \cite{Catalan-Torrecilla2017}). Similarly, the shapes of circular velocity curves (CVCs), reflecting the underlying gravitational potential, central mass distribution, and rotation shear, have been associated with galaxy quenching and dynamical transformation (\cite{Gensior2020}, \cite{Kalinova2022}). Integral-field studies have further revealed that quenching often proceeds inside-out, with star formation suppressed first in central regions, supporting a connection between structural evolution and star formation cessation  (\cite{Sanchez2018}, \cite{Belfiore2018}, \cite{Kalinova2021}). However, these processes have typically been studied separately, leaving open questions about how morphology, dynamics, and quenching interrelate across the full galaxy population.

In this study, we address this question by examining the SFR–$M_*$ relation simultaneously across three complementary galaxy classification schemes. The Hubble sequence (\cite{Hubble1926}, \cite{Hubble1936}) provides insight into structural evolution. The CVC classification based on the Principal Component Analysis (PCA; \cite{Pearson1901}, \cite{Hotelling1933}) probes the total gravitational potential and central mass concentration of the galaxies  (PCA-CVC classification; \cite{Kalinova2017b}). The quenching classification (\cite{Kalinova2021}) separates galaxies by the spatial distribution of star formation, distinguishing actively star-forming, transitioning, and fully quenched systems. By comparing scaling relations across these schemes, we aim to uncover whether structural, dynamical, or quenching-based properties offer distinct or overlapping perspectives on galaxy evolution (e.g. \cite{Kodaira2023}).

The use of integral-field spectroscopic data from the CALIFA survey (\cite{Sanchez2012}) allows us to study galaxies across the full Hubble sequence -- from ellipticals to late-type spirals—on an equal footing. This is particularly important because molecular and neutral gas tracers are often limited or absent in galaxies towards quenching or fully retired systems (\cite{Noordermeer2007}, \cite{Weijmans2008}, \cite{Bolatto2017}), whereas stellar and ionized-gas kinematics provide a uniform and reliable measure of the underlying dynamical structure and star formation, respectively. Combining these structural, dynamical, and star-formation diagnostics allows us to place the SFR$-M_\ast$ relation within a broader evolutionary framework.

The structure of this paper is as follows. In Section \ref{S:data-analysis} we describe the selected sample and analysis methods. Section \ref{S:results} presents our results, Section \ref{S:discussion} places these findings in context, and Section \ref{S:conclusion} summarizes our conclusions.

\section{Data \& Analysis}
\label{S:data-analysis}
\subsection{Sample properties} 
\begin{figure*}
\includegraphics[width=1\textwidth]{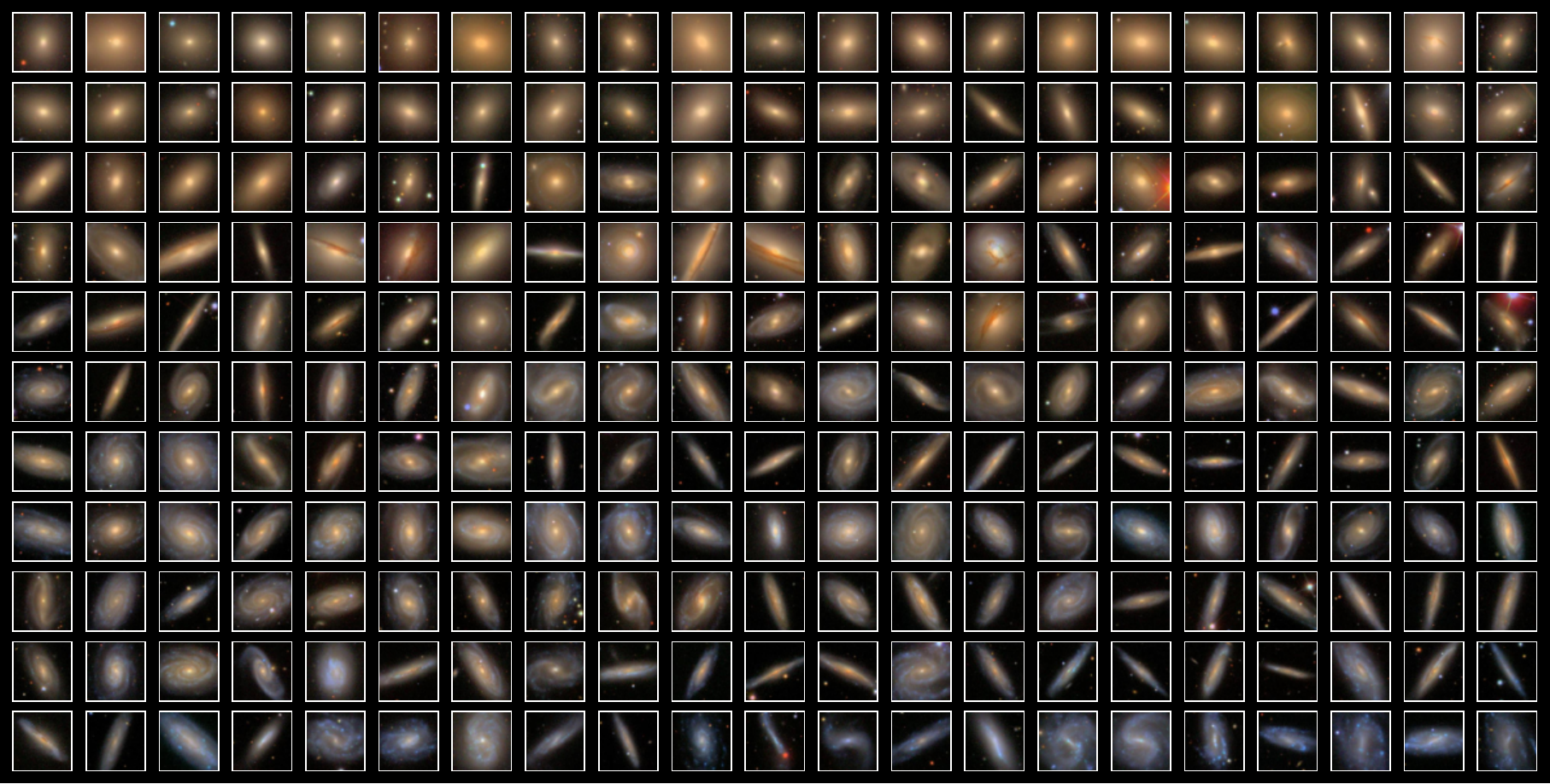}
\caption{\footnotesize{A compilation of individual SDSS DR7 (\cite{Abazajian2009}) multi-color post stamp images of our sample of 231 galaxies, and arranged by Hubble type from elliptical (E1) to late-type spirals (Sdm).}}
\label{fig:sdss-sample}
\end{figure*}

Our exploratory study uses a sample of 231 nearby galaxies (redshifts  $0.005<z<0.03$) from the CALIFA survey (\cite{Sanchez2012}), spanning a wide range of morphologies (from elliptical E1 to late-type spirals Sdm) and $M_\ast$ (from 6$\times$$10^8$ M$_{\odot}$ to 5$\times$$10^{11}$ M$_{\odot}$). This sample is a subset of the the main sample of 238 galaxies analysed in \cite{Kalinova2017b} and  \cite{Kalinova2021} for which both SFR and $M_\ast$ were available. On the other hand, the 238 galaxy sample is a representative subset of the CALIFA mother sample (\cite{Walcher2014}), which characterises the nearby Universe. Further details on the 231 galaxies analysed here can be found in \cite{Kalinova2017b} and  \cite{Kalinova2021}. A compilation of our sample across morphological types is shown in Fig. \ref{fig:sdss-sample}, using multi-color post stamp images from the Sloan Digital Sky Survey (SDSS; Data Release 7, \cite{Abazajian2009}), retrieved from the CALIFA website\footnote{\url{https://califa.caha.es/}}

To study the scaling relations of our sample across Hubble sequence, we use the morphological types defined in \cite{Walcher2014}. These values were collected for the main sample of 238 galaxies by \cite{Kalinova2017b} (see their Table B1), and adopted here throughout the analysis of this study. The stellar mass and SFR used in this study were calculated by \cite{Kalinova2021} using the stellar population synthesis (SPS) method. Further details about the SPS analysis can be found in related works of \cite{Cid-Fernandes2013}, \cite{Gonzalez-Delgado2015}, \cite{deAmorim2017}, \cite{Garcia-Benito2017}, \cite{Garcia-Benito2019}. The star-formation quenching stage (QS) of the galaxies comes from Table B1 of \cite{Kalinova2021}. 

\subsection{Galaxy classification schemes applied in this study}
For our analysis, we use three complementary galaxy classification schemes for the same sample of galaxies, which target a wide range of galaxy structures, mass distributions, and SFRs.

The Hubble classification (also called the Hubble sequence or tuning-fork diagram; \cite{Hubble1926}, \cite{Hubble1936}) is the first categorization scheme adopted in our study. It classifies galaxies based on their morphology, taking into account their shape and structure. Elliptical galaxies span the E0–E7 classes, where the number indicates the apparent ellipticity, from nearly circular (E0) to highly elongated (E7). Lenticular systems are denoted S0, while spiral galaxies range from Sa to Sdm, with earlier types (Sa–Sb) exhibiting larger bulges and tightly wound arms, and later types (Sc–Sdm) showing smaller bulges and more open spiral structures. Irregular galaxies (Irr) include those with disturbed or asymmetric morphologies. Our studied sample, which consists of 231 galaxies, covers morphologies from E1 to Sdm. Although the Hubble classification distinguishes between barred and unbarred spirals, this study does not explore the influence of bars on the scaling relations; this will be addressed in subsequent studies.

\begin{figure*}
\includegraphics[width=1\textwidth]{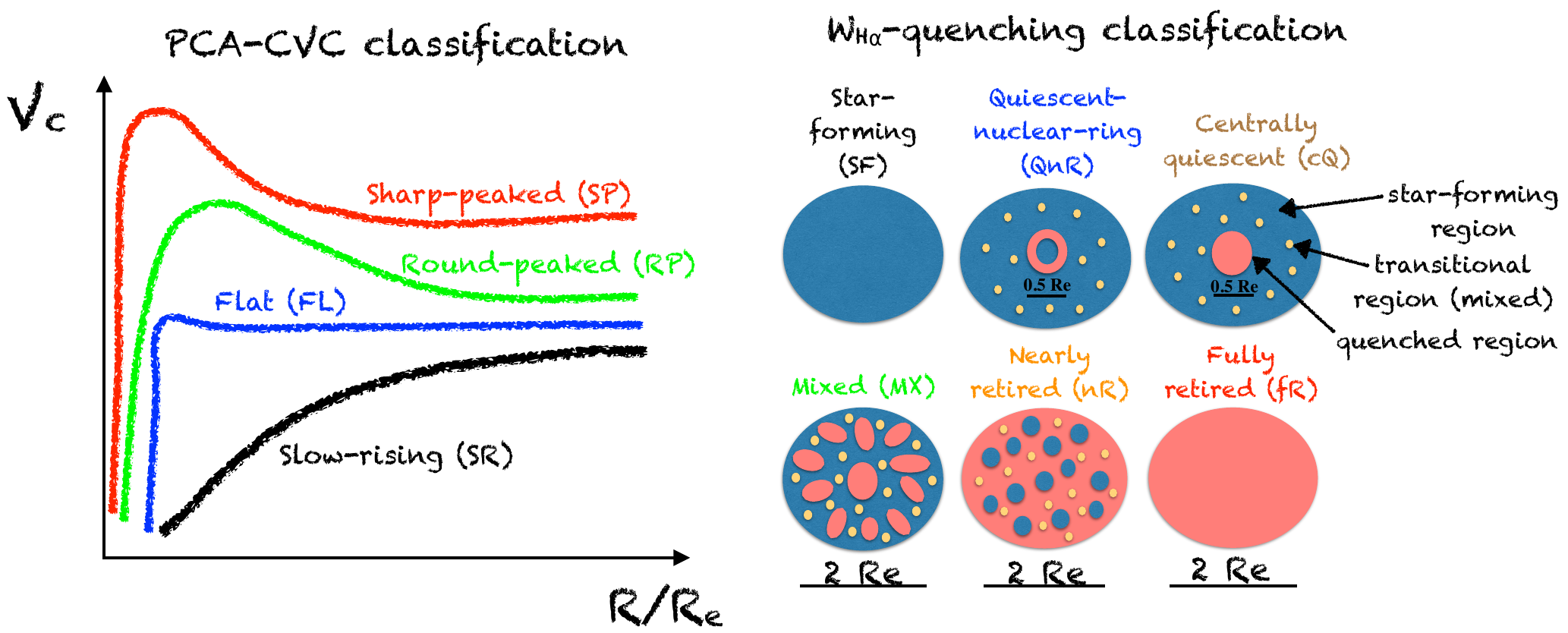}
\caption{\footnotesize{Sketches of two of the three classification schemes used in this study. \emph{Left:} PCA--CVC classification \cite{Kalinova2017b}, defining four galaxy classes with different CVC shapes (SR, FL, RP, and SP). \emph{Right:} W$_\mathrm{H\alpha}$--quenching classification (\cite{Kalinova2021}), distinguishing six classes based on characteristic patterns of ionised gas traced by W$_\mathrm{H\alpha}$ maps (SF, QnR, cQ, MX, nR, and fR)}.}
\label{fig:sketches}
\end{figure*}

The second classification adopted in our analysis is based on the work of \cite{Kalinova2017b}, shown in the left panel of Fig. \ref{fig:sketches}. The authors investigate the shape and amplitude of gravitational potential of the galaxies using unsupervised machine learning techniques. They apply PCA and k-means clustering to classify 238  galaxies according to their CVCs. They define four main CVC classes, which described a changing shape and amplitude of the studied curves: slow-rising (SR), flat (FL), round-peaked (RP), and sharp-peaked (SP). The left panel of Fig. \ref{fig:sketches} shows a sketch of the PCA-CVC classification.

The third classification is based on the recent study of \cite{Kalinova2021}, who investigate 238 nearby CALIFA galaxies, and propose a two--dimensional emission line classification, called  QuestNA (QUEnching STages and Nuclear Activity; see the right panel of Fig. \ref{fig:sketches}). It categorises galaxies based on their ionised gas distribution, marking peculiar patterns in their CALIFA W$_{\rm{H_{\alpha}}}$ maps (obtained through Pipe3D pipeline; \cite{Sanchez2016c},\cite{Sanchez2016b}), and  distinguishes whether the galaxies are  active (i.e., containing weak or strong active galactic nucleus) or non-active.
If galaxies are examined solely based on their QSs, defined by their ionized gas distribution patterns, the authors refer to the QuestNA categorization as ``quenching classification''. Similarly, when galaxies are examined based on their nuclear activity, the classification is referred to as ``nuclear activity classification''. Given the small number of active galaxies in our sample, we focus on investigating the scaling relations according to their ionised gas patterns (i.e., using the quenching classification).

The quenching classification consists of six main QSs, which represent distinct spatial patterns between quenching and star-forming regions, and are defined 
by W$_{\rm{H_{\alpha}}}$ thresholds established in  \cite{Kalinova2021}. In this framework, regions within the field of view of each galaxy can be classified as star-forming (W$_{\rm{H_{\alpha}}} > $ 6 {\AA}), mixed  (3 $<$ W$_{\rm{H_{\alpha}}} \leq $ 6 {\AA}), or retired (W$_{\rm{H_{\alpha}}} \leq $ 3 {\AA}); see also \cite{CidFernandes2011} and \cite{Sanchez2014}. 
Given these criteria, the authors identify six QSs as follows: star-forming (SF; fully dominated by recent star-formation), quiescent-nuclear-ring (QnR; presence of a quiescent-ring structure in the central regions, but still dominated by star formation in the outer-skirts), centrally quiescent (cQ; quiescent inner region within 0.5 effective radius of the galaxy), mixed (MX; no clear patterns in the ionised gas distributions), nearly retired (nR; quiescent galaxies with little star formation regions) and fully retired (fR; completely quiescent object up to two effective radii). 

\section{Results}
\label{S:results}
Using the galaxy classification schemes described above, we investigate the connection between internal structure, gravitational potential, and star-formation quenching of galaxies in galaxy evolution context by examining key classical scaling relations, such as the SFR$-M_\ast$ diagram.

\subsection{SFR$-M_\ast$ diagrams across galaxy classification schemes} 
\begin{figure*}
\centering
\includegraphics[width=0.6\textwidth]{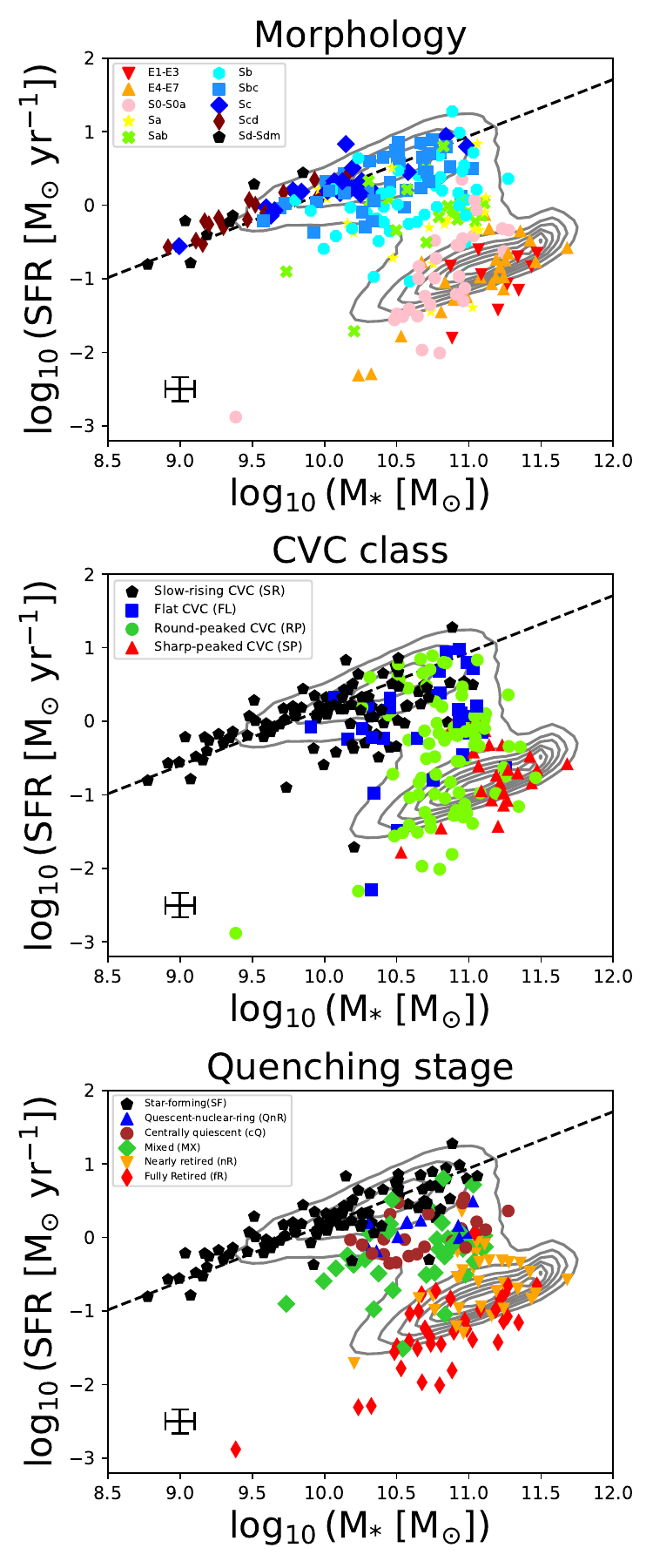}
\caption{\footnotesize{SFR$-M_\ast$ diagram of the studied sample across galaxy classification schemes. The dashed line represents the model of the star-forming MS by \cite{Elbaz2007}, while the error bars -- the typical uncertainties of the $M_\ast$ and the SFR, adopted from \cite{CidFernandes2014} and \cite{Gonzalez-Delgado2017}, respectively. Late-type galaxies follow the star-forming MS, while earlier types, with RP-CVCs and SP-CVCs, and advanced QSs occupy progressively lower SFRs.} The panels together highlight the consistent link between galaxy structure, dynamics, and the suppression of star formation. }
\label{fig:SFRMst}
\end{figure*}

The SFR$-M_\ast$ diagram across the Hubble sequence has been extensively investigated in the literature (e.g. \cite{Elbaz2007}, \cite{Gonzalez-Delgado2016}, \cite{Sanchez2018}, \cite{Bluck2019}). In the top panel of Fig. \ref{fig:SFRMst}, we present the distribution of our galaxy sample across morphological types on the SFR–$M_{\mathrm{*}}$ plane. Late-type galaxies (Sd–Sdm, Scd, and Sc) lie along the star-forming MS, consistent with ongoing and widespread star formation. The deviation from the MS begins around morphological type Sbc, whose galaxies extend from the MS toward the so-called ``blue cloud'', indicating the onset of quenching processes. Progressively earlier morphologies (E, S0--S0a, and some of Sa types) occupy the region below the MS, corresponding to the ``green valley'' and ``red sequence'', where star formation is largely suppressed. This morphological stratification reflects the well-known connection between structural transformation and the decline of star-forming activity (e.g. \cite{Liu2019},\cite{deSa-Freitas2022}).

For the purposes of this study, we further examine how the shapes of the CVCs that traces the total gravitational potential of the galaxies(\cite{Kalinova2017b}) vary across the SFR–$M_{\mathrm{*}}$ diagram (middle panel of Fig. \ref{fig:SFRMst}). The CVC shape serves as a proxy for the internal mass distribution and central concentration of a galaxy. A clear CVC sequence emerges: SR CVCs populate the star-forming MS, representing dynamically cold, disk-dominated systems; FL CVCs deviate slightly (by $\sim$0.5 dex) from the MS but remain actively star-forming; RP CVCs primarily occupy the transition region of the green valley, connecting the high-mass end of the MS with the lower-mass end of the red sequence; and SP CVCs are found almost exclusively at the high-mass end of the red sequence, where galaxies are largely retired. This progression suggests that increasing central mass concentration is tightly linked to the suppression of star formation—from the blue cloud, through the green valley, to the red sequence.

The third panel\footnote{This panel, along with the related third panel of Fig. \ref{fig:dSFR}, was already presented in \cite{Kalinova2021}; we include it here for the purpose of a consistent comparison of the SFR$-M_\ast$ diagram} across different galaxy classification schemes. of Fig. \ref{fig:SFRMst} shows the SFR versus $M_\ast$ across different QSs.  It illustrates that galaxies in the early QS (SF and QnR) generally occupy the star-forming MS, with relatively high SFRs at a given $M_\ast$. As galaxies progress through the intermediate QS (cQ and MX), their global SFRs gradually decrease, moving below the MS. Galaxies in the late QS (nR and fR) populate the high-mass, low-SFR region, characteristic of the red sequence. This distribution highlights the progressive suppression of star formation as galaxies evolve, with the quenching process occurring more rapidly in higher-mass systems ($\log$ M$_* >$  10.3 M$_{\odot}$) .

Furthermore, it is noteworthy that the distribution of CVC classes on the SFR$-M_\ast$ diagram closely mirrors the trends observed for the stellar bulge-to-total mass ratio ($B/T)_\ast$ in local ($z \sim 0.1$) SDSS galaxies (see Fig. 1 in \cite{Bluck2019}). Specifically, SR and FL CVC galaxies correspond to low $(B/T)_\ast$ values, RP CVC galaxies to intermediate $(B/T)_\ast$, and SP CVC galaxies to high $(B/T)_\ast$ ratios. The observed decline in star formation thus appears to be intrinsically linked to the build-up of central bulges or spheroids (e.g. \cite{Catalan-Torrecilla2017}; \cite{Sanchez2018}; \cite{Bluck2019}) and correlates with the depth of the total gravitational potential (e.g. \cite{Gensior2020}).

Together, the morphological, dynamical, and quenching classifications in Fig. \ref{fig:SFRMst} reveal a consistent picture of galaxy evolution: as galaxies grow in mass and develop more centrally concentrated potentials, their star formation is progressively quenched, transforming them from extended star-forming disks into bulge-dominated, dynamically hot systems.

\subsection{Distance from the Main sequence of star formation}
\begin{figure*}
\centering
\includegraphics[width=0.8\textwidth]{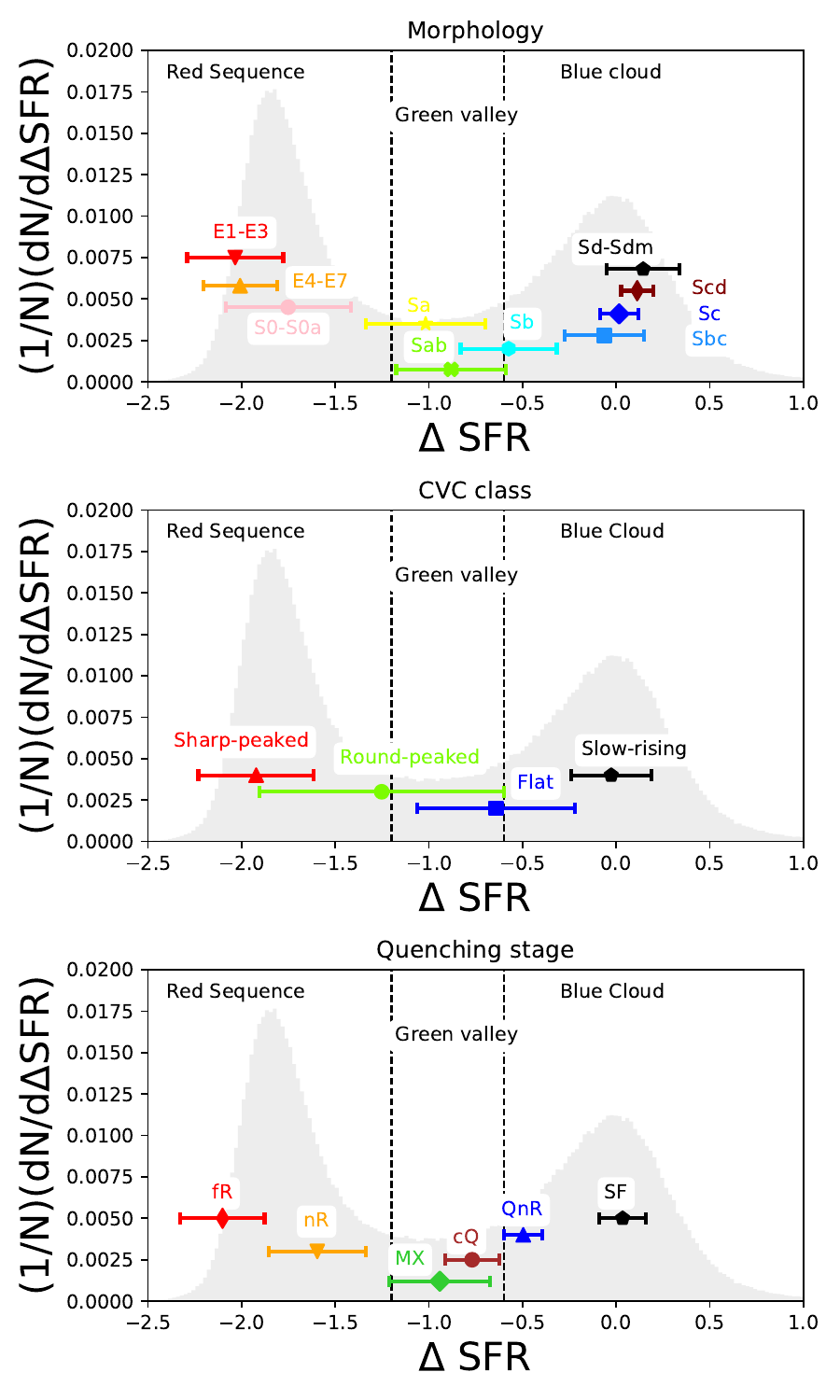}
\caption{\footnotesize{Distance from the MS of the studied sample across different galaxy classification schemes.
Normalised distributions of $\Delta$SFR $\equiv \log(\mathrm{SFR}) - \log(\mathrm{SFR}_{\mathrm{MS}})$ for local SDSS galaxies (grey) and for our sample, separated by morphology (top), CVC class (middle), and QS (bottom).
Vertical dotted lines mark the green-valley region ($-1.2 < \Delta$SFR $< -0.6$) following \cite{Bluck2016}.
We use an arbitrary y-axis for our sample since it is far less numerous than the SDSS galaxy population.}}
\label{fig:dSFR}
\end{figure*}

In Fig. \ref{fig:dSFR}, we summarise the distribution of our galaxies relative to the star-forming MS by means of the logarithmic offset $\Delta$SFR $\equiv \log(\mathrm{SFR}) - \log(\mathrm{SFR}_{\mathrm{MS}})$, as defined by \cite{Elbaz2007}. The quantity $\Delta$SFR is a widely used diagnostic of galaxy evolution, tracing the position of galaxies across the blue cloud, green valley, and red sequence, and therefore serves as a sensitive indicator of star formation quenching (e.g. \cite{Ellison2018}, \cite{Thorp2019}, \cite{Bluck2020b}, \cite{Colombo2020}). In particular, this method allows us to compare how galaxies in different classes are distributed relative to these three regions.

The y-axis represents the normalised histogram of local SDSS DR7 galaxies at $z \sim 0.1$, binned by $\Delta$SFR. For reference, we adopt the division from \cite{Bluck2016}, where $-1.2 < \Delta$SFR $< -0.6$ defines the green valley, separating the blue cloud (actively star-forming galaxies) from the red sequence (quenched systems). The coloured symbols mark the median $\Delta$SFR values for each morphological or CVC class, with the error bars showing the median absolute deviation. Since our sample is much smaller than the SDSS reference population, our histograms use an arbitrary vertical scaling.

The top panel of Fig. \ref{fig:dSFR} clearly shows a continuous morphological sequence along the $\Delta$SFR axis. Late-type spirals (Sc–Sd) dominate the blue cloud, consistent with their high specific SFRs and gas-rich, disk-dominated structures. Intermediate spirals (Sb–Sbc) span the upper portion of the green valley, reflecting the onset of quenching as gas reservoirs are gradually depleted or stabilised. Early-type spirals (Sa–Sab) and lenticulars (S0–S0a) populate the lower green valley and approach the red sequence, indicating a more advanced stage of star formation suppression. Finally, elliptical galaxies (E1--E7) are concentrated entirely in the red sequence, consistent with fully quenched stellar populations.

This morphological stratification mirrors the canonical Hubble sequence of structural evolution, connecting disk-dominated, star-forming galaxies to bulge-dominated, passive systems. The smooth transition in $\Delta$SFR across morphological types supports the view that galaxy quenching proceeds gradually, in tandem with the growth of central bulges and the transformation of kinematic structure.

A parallel sequence is seen for the CVC classes (middle panel). SR and FL CVC galaxies primarily occupy the blue cloud, consistent with their extended, rotation-dominated disks and low central mass concentrations. RP systems populate the green valley, covering the transition from active to quiescent regimes. These galaxies likely represent systems undergoing structural reconfiguration—where central mass build-up (from bars, bulges, or gas inflows) begins to stabilise the disk against further star formation (e.g. \cite{Martig2009}, \cite{Gensior2020}). Finally, SP CVC galaxies are confined to the red sequence, consistent with highly concentrated gravitational potentials and fully quenched stellar populations.

The amplitude and steepness of the CVCs increase systematically with the distance from the MS, reinforcing the physical link between a galaxy’s central potential shape and its current SFR. As galaxies evolve, the deepening of their potential wells—through secular evolution or merger-driven processes—appears to lead to gas stabilisation and the cessation of star formation (e.g. \cite{Gensior2020}). The overlap between RP CVCs and intermediate morphologies (Sa–-Sb) across the green valley suggests that these systems represent the key transitional phase of quenching.

Together, the morphological and CVC distributions in Fig. \ref{fig:dSFR} outline a coherent evolutionary pathway. Late-type, SR galaxies in the blue cloud evolve toward RP and SP systems as their morphology and mass profiles become increasingly bulge-dominated. This transformation is accompanied by a steady decline in $\Delta$SFR, marking the progression from active star formation to full quiescence.

The third panel of Fig. \ref{fig:dSFR}, re-plotted from \cite{Kalinova2021}, displays the normalized distributions of $\Delta$SFR (the offset from the star-forming MS) for galaxies in different QSs. We include it here to ensure a consistent comparison with our analysis of the SFR$-M_\ast$ diagram across QSs. As found in \cite{Kalinova2021}, SF galaxies cluster around $\Delta$SFR $\approx$ 0, consistent with active star formation along the MS. QnR and cQ galaxies show a moderate shift toward lower $\Delta$SFR values, indicating a decline in central star formation activity. MX systems occupy an intermediate regime, bridging the transition from active to quenched populations. Finally, nR and fR galaxies exhibit the most negative $\Delta$SFR values, well below the MS, consistent with global suppression of star formation. This systematic shift of $\Delta$SFR across QSs demonstrates the progressive depletion of star formation as galaxies evolve from the star-forming sequence to the red sequence.

Overall, Fig. \ref{fig:dSFR} provides strong evidence that morphological transformation and dynamical evolution are tightly coupled with star formation quenching. The increasing central mass concentration--whether traced by morphology, CVC shape, or QS--appears to play a fundamental role in regulating a galaxy’s position relative to the MS and, ultimately, its evolutionary path.

\section{Discussion}
\label{S:discussion}
\begin{figure*}[t]
\includegraphics[width=1\textwidth]{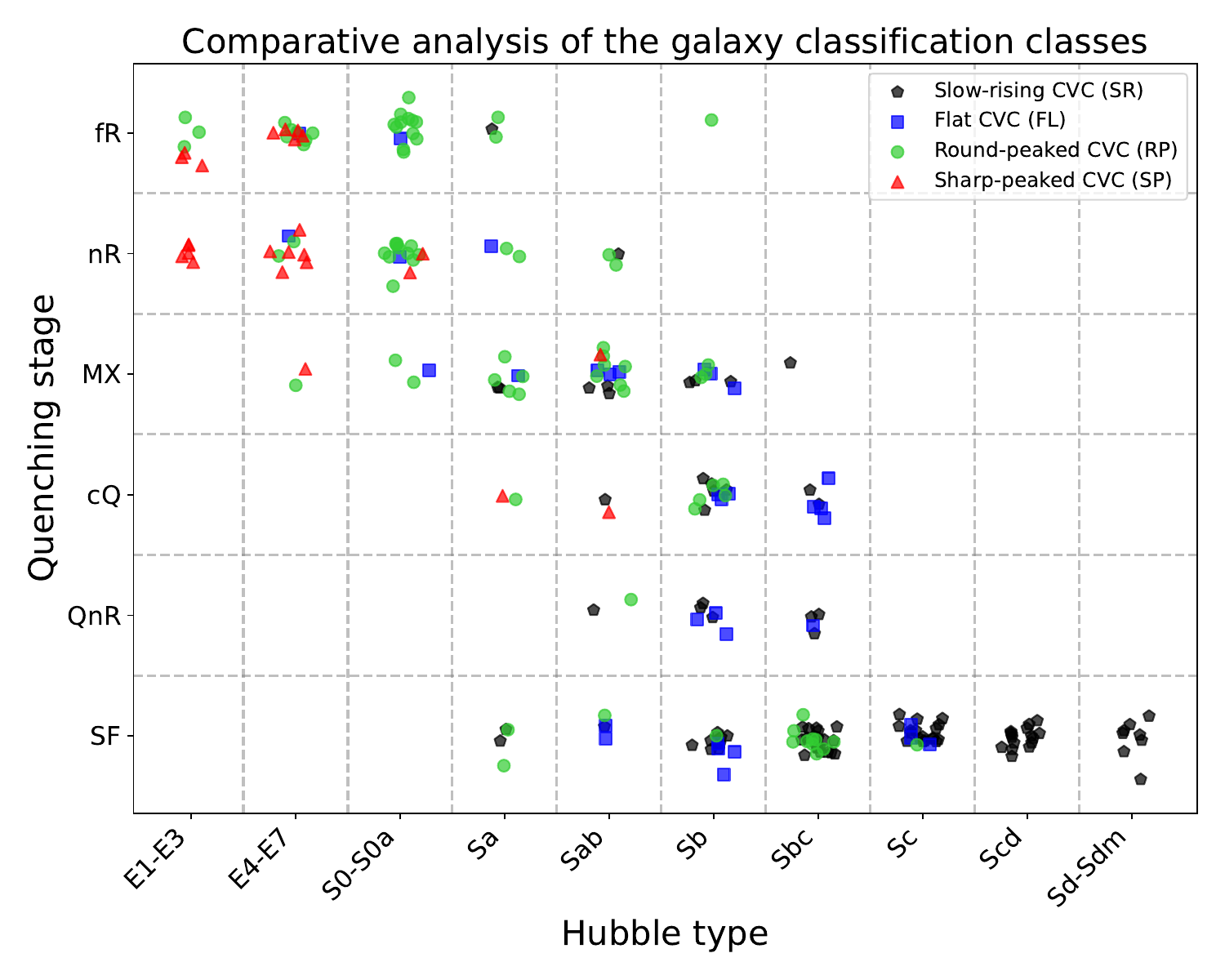}
\caption{\footnotesize{Scatter-grid plot of galaxies by morphology and QS. Each cell shows individual galaxies as points, coloured by CVC class, with the grid layout allowing comparison of distributions across morphology–quenching combinations. A small random shift to the galaxy location is added artificially to avoid large overlaps between the points.}}
\label{fig:3d-scatter}
\end{figure*}

\begin{figure*}[h!]
\centering
\includegraphics[width=0.98\textwidth]{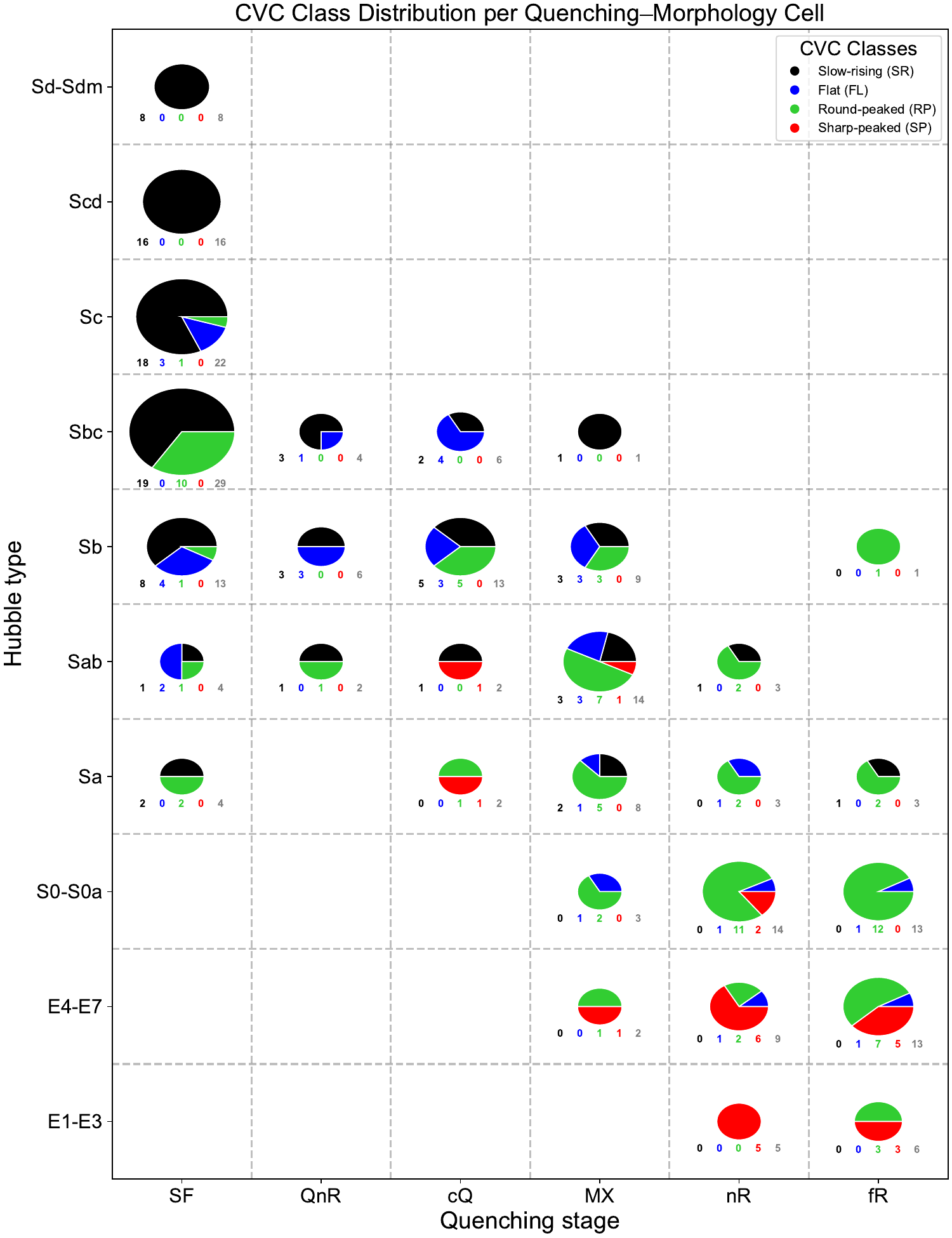}
\caption{\footnotesize{Distribution of CVC classes across galaxy morphology and QS. Each cell corresponds to a specific morphology–quenching combination and contains a proportional pie chart showing the fractional composition of the four CVC classes. The radius of each pie is scaled to the total number of galaxies in that cell, and the numbers below indicate the raw counts per CVC class and the total number of galaxies (in grey).}}
\label{fig:3d-pie}
\end{figure*}

The results presented in Figs.~\ref{fig:SFRMst} and \ref{fig:dSFR} reveal a consistent and interconnected picture of galaxy evolution, in which morphology, internal dynamics, and star formation activity are tightly linked. The position of a galaxy on the SFR$-M_\ast$ diagram, as well as its offset from the star-forming MS ($\Delta$SFR), traces a smooth progression from late-type, disk-dominated, star-forming systems to early-type, bulge-dominated, quenched galaxies (\cite{Brinchmann2004}, \cite{Noeske2007}, \cite{Whitaker2012}, \cite{Speagle2014}). This trend is mirrored by the sequence of CVC classes, from SR to SP, indicating that the build-up of central mass and the deepening of the gravitational potential are closely connected to the suppression of star formation (\cite{Gensior2020}, \cite{Bluck2019}, \cite{DekelBurkert2014}).

The morphological distributions across $\Delta$SFR highlight a parallel structural trend: late-type spirals (Sc–-Sd) occupy the blue cloud, intermediate spirals (Sa–-Sb) populate the green valley, and early-type systems (E--S0) dominate the red sequence. This smooth progression supports the long-established view that quenching is accompanied by morphological transformation and bulge growth, either through internal secular evolution such as bar-driven inflows and disk instabilities (\cite{Kormendy2004}, \cite{Athanassoula2013}, \cite{Spinoso2017}) or through merger-driven processes (\cite{Hopkins2008}, \cite{Naab2014}). The overlap between RP CVCs and intermediate morphologies in the green valley further suggests that this stage represents a key transitional phase, where central mass accumulation and star formation suppression occur simultaneously (\cite{Martig2009}, \cite{Martig2013}, \cite{Tacchella2016}).

Figures~\ref{fig:3d-scatter} and \ref{fig:3d-pie} provide a more detailed view of these trends. The scatter-grid plot shows how galaxies populate the combined morphology–quenching parameter space, while the pie-grid visualization highlights systematic changes in the distribution of CVC classes. Actively star-forming galaxies with low central concentration (SR or FL CVCs) dominate the blue cloud, while highly concentrated, bulge-dominated systems (RP and SP CVCs) dominate the red sequence. Intermediate systems occupy the green valley and display a wide range of CVC shapes and QSs, emphasizing that the transition from active to quiescent states is continuous and multifaceted rather than discrete. These findings are consistent with results from other large IFU surveys, such as SAMI and MaNGA, which also reveal a tight connection between morphology, internal kinematics, and star formation efficiency (\cite{Cappellari2016}, \cite{Bluck2020b}, \cite{Brownson2022}).

The spread of CVCs and QSs among intermediate morphologies (Sa--Sbc) suggests that while bulge growth and morphological stabilization play a major role in quenching, they are not the only mechanisms at work. Other processes -- such as bar-driven inflows, disk instabilities, mergers, AGNs (active galactic nuclei) or stellar feedback, and environmental effects—can contribute to gas depletion and star formation suppression (\cite{Croton2006}, \cite{Hopkins2008}, \cite{Peng2010}, \cite{Naab2014}, \cite{Terrazas2020}). The combination of these internal and external factors likely operates with varying efficiencies and timescales, producing the observed diversity among galaxies in the green valley.

Examining $\Delta$SFR strengthens this interpretation. Galaxies with higher central mass concentrations systematically fall below the MS, reflecting a progressive decline in star formation as they evolve from the blue cloud through the green valley to the red sequence. The alignment of the CVC class with the bulge-to-total stellar mass ratio, as reported in \cite{Bluck2019} and \cite{Bluck2022}, underscores the role of central potential depth as a key regulator of star formation. RP CVC systems in the green valley, with intermediate potentials and quenching fractions, exemplify the transitional phase where partial stabilization and gas depletion are already under way.

Altogether, the combined trends of morphology, QS, and CVC class provide a coherent and physically motivated framework for understanding galaxy evolution. Slow-rising and FL CVCs correspond to dynamically cold, gas-rich disks that sustain high star formation rates; RP-CVC systems trace the onset of central concentration and partial stabilization; and SP-CVC systems represent dynamically hot, gas-poor, quenched spheroids. The tight connection between dynamical state and star formation activity suggests that internal structure and gravitational potential depth might be the primary regulators of quenching, with additional contributions from external or feedback-related processes.

\section{Concluding remarks}
\label{S:conclusion}
We have investigated the relationship between SFR and $M_\ast$ across three complementary galaxy classification schemes —morphology, CVC shape, and QS — to explore how galaxy structure and dynamics regulate star formation. Our results reveal a continuous evolutionary sequence connecting disk-dominated, star-forming galaxies to bulge-dominated, quenched systems, accompanied by systematic changes in central mass concentration and gravitational potential depth.

Late-type spirals (Sc--Sd) dominate the star-forming MS with SR or FL CVCs, while early-type galaxies (E--S0) occupy the red sequence with SP CVCs and fully quenched stellar populations. Intermediate spirals (Sa--Sc) span a wide range of CVC shapes and QSs, highlighting the green valley as a transitional regime. This diversity indicates that quenching in these systems is not driven by a single mechanism; rather, multiple processes—including bulge growth, bar-driven inflows, disk instabilities, feedback, and environmental effects—likely act together or sequentially to suppress star formation.

The $\Delta$SFR distributions reinforce this picture: galaxies with higher central mass concentrations systematically fall below the MS, reflecting declining star-formation efficiency. The alignment of CVC shapes, bulge-to-total mass ratios, and QSs indicates that the build-up of central mass and the deepening of the gravitational potential are key regulators of star formation, while intermediate morphologies most likely represent the complex, multi-channel nature of galaxy quenching.

Overall, our findings support a coherent, gradual, and interconnected view of galaxy evolution, in which internal structure, dynamics, and star-formation activity evolve together. The observed trends across morphology, CVC class, and QS provide a physically motivated framework for understanding transitions from the blue cloud to the red sequence.

In subsequent papers, we will expand on these connections, quantifying the relative importance of structural and dynamical factors in star-formation quenching. In particular, we will explore the interplay between galaxy structure, gas fraction, and the evolution of central mass concentration to build a comprehensive, quantitative framework for understanding how internal dynamics and baryonic composition govern the cessation of star formation.

\section{Acknowledgements}
\footnotesize{We thank the anonymous referee for the helpful comments that improved the quality of this manuscript.
DC gratefully acknowledges the Collaborative Research
Center 1601 (SFB 1601 sub-project B3) funded by the Deutsche Forschungsgemeinschaft (DFG, German Research Foundation) –
500700252.
In this study, we made use of the data of the first legacy survey, 
the Calar Alto Legacy Integral Field Area (CALIFA) survey, 
based on observations made at the Centro Astron\'omico
Hispano Alem\'an (CAHA) at Calar Alto, operated jointly by the Max Planck-Institut
f\"ur Astronomie and the Instituto de Astrof\'isica de Andaluc\'ia
(CSIC).
Funding for the Sloan Digital Sky Survey IV has been provided by
the Alfred P. Sloan Foundation, the U.S. Department of Energy Office of
Science, and the Participating Institutions. SDSS-IV acknowledges
support and resources from the Center for High-Performance Computing at
the University of Utah. The SDSS web site is www.sdss.org.
SDSS-IV is managed by the Astrophysical Research Consortium for the 
Participating Institutions of the SDSS Collaboration including the 
Brazilian Participation Group, the Carnegie Institution for Science, 
Carnegie Mellon University, the Chilean Participation Group, the French Participation Group, Harvard-Smithsonian Center for Astrophysics, 
Instituto de Astrof\'isica de Canarias, The Johns Hopkins University, 
Kavli Institute for the Physics and Mathematics of the Universe (IPMU) / 
University of Tokyo, Lawrence Berkeley National Laboratory, 
Leibniz Institut f\"ur Astrophysik Potsdam (AIP),  
Max-Planck-Institut f\"ur Astronomie (MPIA Heidelberg), 
Max-Planck-Institut f\"ur Astrophysik (MPA Garching), 
Max-Planck-Institut f\"ur Extraterrestrische Physik (MPE), 
National Astronomical Observatory of China, New Mexico State University, 
New York University, University of Notre Dame, 
Observat\'ario Nacional / MCTI, The Ohio State University, 
Pennsylvania State University, Shanghai Astronomical Observatory, 
United Kingdom Participation Group,
Universidad Nacional Aut\'onoma de M\'exico, University of Arizona, 
University of Colorado Boulder, University of Oxford, University of Portsmouth, 
University of Utah, University of Virginia, University of Washington, University of Wisconsin, 
Vanderbilt University, and Yale University.
This research made use of the open-source python packages as \texttt{Astropy} (\cite{Price2018astropy}), \texttt{SciPy} (\cite{SciPy2020}), \texttt{NumPy} (\cite{Harris2020NumPy}), and \texttt{Matplotlib} (\cite{Hunter2007matplotlib}). }


\bibliographystyle{apalike}
\bibliography{BIB_v10}


\end{document}